\documentstyle[12pt]{article}

\def\a{\alpha}
\def\o{\omega}
\def\b{\beta}
\def\g{\gamma}
\def\t{\cal T}
\def\lt{\langle}
\def\rt{\rangle}
\def\f{\phi}
\def\P{\Phi}
\def\cg{\cal G}

\newcommand{\be}{\begin{equation}}
\newcommand{\ee}{\end{equation}}

\begin{document}
\baselineskip= 24 truept
\begin{titlepage}

\title { Topological Kac-Moody Algebra and  Wakimoto Representation}

\author{Abbas Ali$^1$ and Alok Kumar$^2$\\
Institute of Physics, Bhubaneswar-751005, INDIA.}

\footnotetext[1]{e-mail:abbas\%iopb@shakti.ernet.in}
\footnotetext[2]{e-mail:kumar\%iopb@shakti.ernet.in}

\date{}
\maketitle

\thispagestyle{empty}

\begin{abstract}

It is shown, using the Wakimoto representation, that the level zero SU(2)
Kac-Moody conformal field theory is topological and can be obtained
by twisting an  N=2 superconformal  theory. Expressions for the associated
N=2 superconformal generators are  written down and the Kac-Moody generators
are shown to be BRST exact .

\end{abstract}

\bigskip
\vfil

\rightline{IP/BBSR/91-48}

\end{titlepage}

\eject

Recently there has been a great deal of interest
in topological field theories \cite{wit}-\cite{hl}.
These theories are characterised by
the existence of a BRST charge such that the energy momentum tensor
of the theory is  BRST exact \cite{wit}. Two dimensional version of the
theories,{\it e.g.}, the topological conformal field theories (TCFT's)
have also been studied extensively \cite{ey}\cite{dvv}\cite{kl}
because of their connection to two dimensional quantum gravity
and matrix models\cite{dwit}.
 Supersymmetric
as well as higher genus generalization of TCFT's have also been
reported \cite{hl}-\cite{akms}.

In an illuminating paper, the topological nature of a class  of $c = 0$
conformal field theories (CFT's) was
shown by Eguchi and Yang\cite{ey}. This result was proved
using the free boson, {\it viz.}, Feigin-Fuchs
realization of the CFT's. Topological nature of $c = 0$ theories have
also been shown for $W_N$ theories \cite{lerche} as well as for
 a class of coset models \cite{ehy}.

In this note, we extend the results of \cite{ey} to level zero
$SU(2)$ Kac-Moody Algebra. We show, using the Wakimoto realization of
$SU(2)$ Kac-Moody and associated conformal algebra, that this theory
is topological. We now start by
reviewing the essential features of \cite{ey}. In Feigin-Fuchs construction,
the expression for the energy momentum tensor ${\t}(z)$ is

\be
{\t}(z) = -{1\over 2}( \partial {\f (z)})^2 + i\a_0\partial^2\f (z)
\ee
where $\f (z)$ is a free boson with the two point function:
\be
\lt\f (z) \f (\o) \rt = - log(z-\o).
\ee

The central charge of the theory, $c = 1 - 12 \a_0^2 $, vanishes for
$\a_0 = {1\over 2\sqrt{3}}$.  The screening charge of the Feigin-Fuchs
construction,$Q = \oint G(z) dz  = \oint e^{i \sqrt{3} \f (z)}dz$,
satisfies the nilpotency condition
$Q^2 = 0$. Moreover, using the operator product expansions (OPE's)
it can be shown that for $\a_0 = {1\over 2\sqrt{3}}$

\be
\{Q, \bar {\cg}(z)\} = {\t}(z)
\ee
where
\be
\bar {\cg}(z) = e^{-i \sqrt{3}\f (z)}.
\ee

Therefore if $Q$ is identified as the BRST charge, the energy-momentum
tensor is BRST exact, and the  theory is topological. This topological
field theory is in fact a twisted $N = 2$ superconformal theory.
To see this, we write the energy-momentum tensor as

\be
{\t}(z) = {\t}_{N=2}(z) + {1\over 2}J(z)
\ee
where
\be
{\t}_{N=2}(z) = -{1\over 2}(\partial \f (z))^2
\ee
 and
 \be J(z) = {i\over \sqrt{3}}\partial \f (z).
 \ee

It is now straightforward to show that the operators ${\t}_{N=2}(z)$,
$J(z)$, ${\cg}(z)$, and $\bar {\cg}(z)$ satisfy the $N = 2$ superconformal
algebra with central charge $c = 1$.

We now extend these results to the SU(2) Kac-Moody CFT using the Wakimoto
realization \cite{wak}\cite{dots}. In this realization\cite{dots}, the SU(2)
 generators are written as
\be
J^+(z) = \o ^+(z)
\ee
\be
J^0(z) = -i(\o(z)\o ^+(z) + {1\over {2\a_0}}\partial \f (z))
\ee
\be
J^- = \o(z)\o(z)\o ^+(z) + ik\partial \o (z)+ {1\over {\a_0}}\partial
\f (z)\o (z)
\ee
where $\o (z)$, $\o^+(z)$ and $\f (z)$ are free boson  fields
with the two point functions:
\be
\lt \o (z_1)\o^+(z_2)\rt = -\lt\o^+(z_1)\o (z_2)\rt = {i\over (z_1 - z_2)}
\ee
\be
\lt \f (z_1)\f (z_2)\rt = -log(z_1-z_2).
\ee

In this realization, the level $k$  of the $SU(2)$ Kac-Moody
 algebra \cite{dots} is
\be
k = -2 + {1\over {2\a_0^2}}.
\ee

The Sugawara construction leads to the following
expression for energy-momentum:

\be
T(z) = -{1\over 2}(\partial \f (z))^2 + i\a_0\partial^2\f (z)
+ i\o^+(z)\partial \o (z).
\ee
The corresponding central charge is
\be
c = 3 - 12\a_0^2 = {3k\over {k+2}}.
\ee

We now specialize to the case of level zero Kac-Moody algebras.
In this case, using eqs.(13) and (15), we
get $\a_0 = {1\over 2}$ and $c = 0$. Also, the eqs.(8-10) and (14)
become
\be
J^+(z) = \o ^+(z)
\ee
\be
J^0(z) = -i(\o (z)\o ^+(z) +\partial \f (z))
\ee
\be
J^-(z) = \o (z)\o (z)\o ^+(z) + 2\partial \f (z)\o (z)
\ee
and
\be
T(z) = -{1\over 2}(\partial \f (z))^2 + {i\over 2}\partial^2\f (z) +
i\o^+(z) \partial \o (z)
\ee
respectively. Here we would like to remark that although the algebra
satisfied by the generators (16)-(19) has level $k = 0$, but it is
not a classical algebra.
This is because OPE's of the generators involve multi-contractions.

Now we show that the Kac-Moody CFT represented by eqs.(16-19) is topological
and is a twisted version of an $N = 2$ superconformal theory. To illustrate
the $N = 2$ structure we rewrite
\be
T(z) = T_{N=2}(z) + {1\over 2}J^{U(1)}(z)
\ee
where
\be
J^{U(1)}(z) = i \partial \f (z)
\ee
and
\be
T_{N=2}(z) = -{1\over 2}( \partial \f (z))^2 + i\o^+(z) \partial \o(z).
\ee

The supercharges of the $N = 2$ theory can also be obtained from the
knowledge of the operator content of the level zero Kac-Moody algebra.
The screening operator of the $SU(2)$ Kac-Moody algebra is given by
\cite{dots} $\P_+(z) =  \o^+(z)e^{i\f (z)}$. We identify it with one of the
supercharges:
\be
G(z) = \o^+(z)e^{i\f (z)}.
\ee
It can be shown that the operator
\be
\bar G(z) = 2i\partial\o (z) e^{-i\f (z)}
\ee
is the other supercharge. The operators (21)-(24) satify an
N= 2 superconformal algebra with central charge $c = 3$.

The topological nature of the original $k = 0$ Kac-Moody
theory  can now be shown
in the usual manner \cite{ey}\cite{kl}. The BRST charge, defined by
\be
Q = \oint G(z) dz
\ee
is nilpotent. And using the operator algebra
\be
G(z_1)\bar G(z_2) ={2\over (z_1-z_2)^3}+ {2J(z_2)\over
(z_1-z_2)^2}+{{2T(z_2) + \partial J(z_2)}\over (z_1-z_2)}.
\ee
we have
\be
\{Q, \bar G(z)\} = T(z).
\ee
Therefore $T(z)$ is BRST exact and the theory is topological.

\vfil
\eject

Now, by defining the operators,

\be
j^+(z) = e^{-i\f (z)}
\ee
\be
j^0(z) =-i\o (z) e^{-i\f (z)}
\ee
\be
j^-(z) =\o^2(z) e^{-i\f (z)}
\ee
one obtains, using the OPE's:
\be
\{Q, j^{\pm, 0}(z)\} = J^{\pm, 0}(z).
\ee
Therefore $J^{\pm,0}(z)$ are BRST exact and $j^{\pm,0}(z)$ are their BRST
partners

The primary fields of the $SU(2)$ Kac-Moody CFT in the Wakimoto
representation are given by \cite{dots}
\be
\P^j_m(z) = ( \o (z))^{j-m}e^{-2i \a_0 j \f (z)}, m = -j, ... , j.
\ee
The conformal weight of the primary fields $\P^j_m(z)$ is given by
$\Delta^j_m = {j(j+1)\over {k+2}}$. It is known that due to
unitarity considerations \cite{gwit}, $j$ is restricted to
$0 \leq j \leq {k\over 2}$ and hence in the present case
only $j = 0$ primary state survives
and has weight zero. Moreover, due to BRST exactness of
$T(z)$ and  $J^{\pm, 0}(z)$,
Virasoro as well as Kac-Moody secondary states are absent from the BRST
cohomology.

At this point we would like to remark that in the untwisted N=2
theory with $c = 3$ , {\it i.e.}, $\a_0 = 0$, the SU(2) currents in
eqs.(8)-(10) are not well defined.
This is unlike the case of
superconformal extensions of the TCFT where the extra supersymmetry
generator is well defind in the untwisted theory as well. This enables
one to write the full algebra of the untwisted theory in these
cases as $N > 2 $ extended superconformal algebras \cite{km}\cite{noj}.

Finally, we comment on the relation of our work with ref. \cite{ehy}.
In ref.\cite{ehy} the topological nature of the cosets
${G_0\otimes G_k\over G_k}$ was shown which apparently seems to
contain our case. However, one notices
that the BRST charge in eqs.(5), (12), and (13) of ref.
\cite{ehy} is not well defined for k=0.
But, as we have seen, Wakimoto realization overcomes this difficulty.

To conclude, we have proved that the level zero Kac-Moody theory is
topological. One can possibly generalize our work,
using Wakimoto kind of realization for other groups \cite{gera} as
well as coset constructions. Generalization to super Kac-Moody algebras
using the corresponding free field realization \cite{kimura} should
also be possible.

{\sc NOTE ADDED}: After the completion of this work we received a preprint
\cite{yoshii} where $N = 2$ supersymmetry of
$(b, c)$ and $(\b, \g)$ ghost
system was used to construct a
topological SL(2) Kac-Moody algebra.
However we believe that for generalization to
other groups Wakimoto type of realization realization should
be more useful.

\baselineskip 12pt
\vfil
\eject

\end{document}